%% file: Regge_e.tex
\documentclass[ aps,%
floatfix,%
final,%
notitlepage,%
oneside,%
onecolumn,%
nobibnotes,%
nofootinbib,%
superscriptaddress,%
]%
{revtex4}

\RequirePackage[english]{babel}

\usepackage{epsfig}
\usepackage{axodraw}
\usepackage{graphics}
\usepackage{amssymb}

\newcommand{\Run}[1]{}

\newcommand{\JP}{\psi} 
\newcommand{\epem}{e^+e^-} 

\newcommand{\GeV}{\,\mbox{GeV}} \newcommand{\keV}{\,\mbox{keV}}\newcommand{\MeV}{\,\mbox{MeV}}
\newcommand{\fm}{\,\mbox{fm}}

\newcommand{\beq}{\begin{eqnarray}}\newcommand{\eeq}{\end{eqnarray}}
\newcommand{\beqa}{\begin{eqnarray*}}\newcommand{\eeqa}{\end{eqnarray*}}

\newcommand{\Br}{\mathrm{Br}}

\begin{document}

\title{Systematics of heavy quarkonia from Regge trajectories on $(n,M^2)$ and $(M^2,J)$ planes}

\author{S.S. Gershtein}
\email{Semen.Gershtein@ihep.ru}
\author{A.K. Likhoded}
\email{Likhoded@ihep.ru}
\author{A.V. Luchinsky}
\email{Alexey.Luchinsky@ihep.ru}
\affiliation{Institute of High Energy Physics, Protvino, Russia}
\begin{abstract}
In this paper we show that heavy quarckonium states, similar to light mesons, form Regge trajectories in $(n,M^2)$ and $(M^2,J)$ planes and the slope of these trajectories is independent on the quantum numbers of the mesons. This fact can be useful for the prediction of the masses of heavy quarkonia and the determination of the quantum numbers of the newly discovered  states.
\end{abstract}
\maketitle

\section{Introduction}

It is well known that for wide range of potentials the only singularities of the non-relativistic scattering amplitude are the poles in the the orbital momentum $l$ plane (Regge-poles). The position of this poles depends on the energy, i.e. $l=\alpha(s)$, where $\alpha(s)$ is the Regge trajectory that corresponds to physical particles and resonances. 

The collection of the values of the masses of light mesons $q\bar q$ is described fairly good by the linear Regge trajectories both on $(n,M^2)$ and $(M^2,J)$ planes (here $M$ is the mass of the meson and $n$ and $J$ are its radial quantum number and total spin respectively). The linearity of the dependence
\beqa
n &=& \frac{1}{\mu_{q\bar q}^2}M^2+n_0, \qquad J=\frac{1}{\mu_{q\bar q}^2}M^2+J_0
\eeqa
with the universal slope coefficient $\mu^2\sim 1.2\GeV^2$ is typical for the "string" nature of the mesons. In the series of works \cite{Batunin:1989gd,Ida:1977uy} the attempts were made to describe the relativistic string with fermions at the ends. In the limit of massless quarks the relativistic string model leads to linear Regge trajectories $J=\alpha' M^2+\alpha(0)$. When the quark masses are taken into account the trajectory deviates from linear for lower states. Another reason for such  deviation is the screening of the quark-antiquark potential because of the additional $q\bar q$-pair production. This effect results is important for higher states of the mesons.

The spectrum of light mesons was analyzed in the works \cite{Anisovich:2000kx,Baldicchi:1998gt} and the Regge trajectories for this states were presented. Below we will perform the same analysis for heavy quarkonium states using the whole experimental information available at the moment. Such an analysis could be useful both for the comparison of the experimental data with potential model predictions and for the determination of the parameters of regge trajectories for heavy quarkonium.

\section{Charmonium}

\begin{figure}[h]
\includegraphics{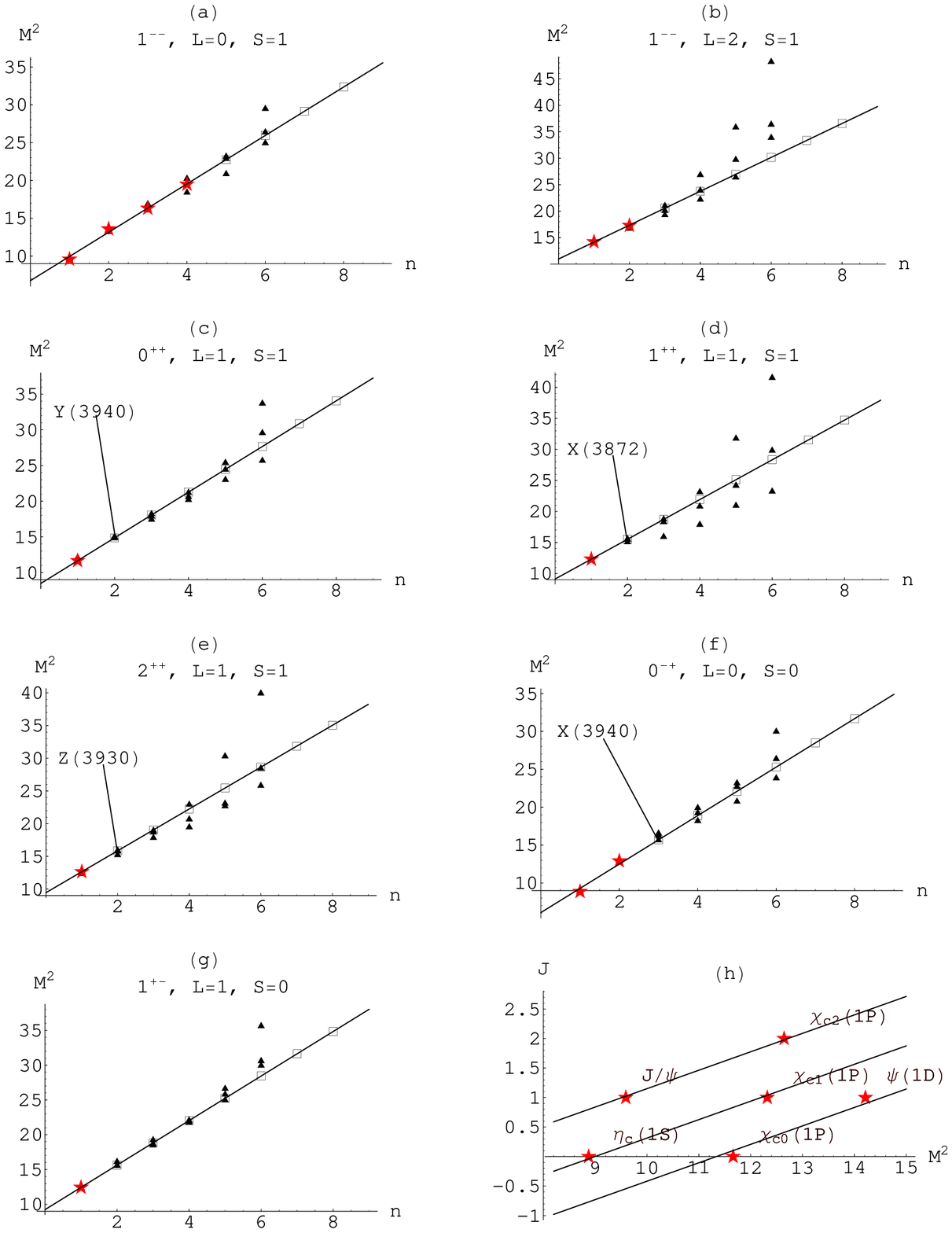}
\caption{
Regge trajectories for charmonium mesons.%
$\bigstar$ --- experimental values, $\blacktriangle$ and $\blacksquare$ --- the results of the works \cite{Anisovich:2005vs} and \cite{Baldicchi:1998gt}, $\square$ --- our predictions
\label{fig:charm}}
\end{figure}

The most extensively studied charmonium mesons are vector states ($J^{PC}=1^{--}$). For this states the value of the total spin of the quark-antiquark pair is $S=1$, while for the orbital momentum of this pair the values $L=0$ and $L=2$ are allowed, so the mixing of the states with different values of $L$ is possible. This effect, however, will not change the masses of the particles significantly and we will neglect it in our paper. On fig.\ref{fig:charm}a,b we show the experimental values of the masses of vector charmonia with $L=0,2$ (red stars) and the predictions presented in the works \cite{Baldicchi:1998gt} and  \cite{Anisovich:2005vs} (they are labeled by the symbols  $\blacksquare$ and $\blacktriangle$ respectively). It is clearly seen that the experimental values are fairly well described by linear Regge trajectories and the mass of the meson $M$ and its radial quantum number $n$ satisfy the relation
\beq
M_n^2 &=& \mu_{c\bar c}^2 n+M_0^2. \label{eq:nM}
\eeq
Here the slope coefficient
\beq
\mu_{c\bar c}^2 &=& 3.2\GeV^2 \label{eq:mucc}
\eeq
is the same both for $L=0$ and $L=2$, and the parameter $M_0$ depends on the value of the orbital momentum. In the table \ref{tab:Charm} we show the values of this parameter, the experimental values of vector meson masses and our predictions for the masses of excited mesons obtained with the help of the formula (\ref{eq:nM}).

The experimental information about the mesons with other values of $L$ and/or $S$ is more poor, since the observation of this states is more complicated. For each of the sets of the values of this quantum numbers the mass of at least one meson is, however, known, so we can determine the parameter $M_0$ in the relation (\ref{eq:nM}) and using a known value of the slope coefficient $\mu^2$ construct the corresponding Regge trajectory. In the table \ref{tab:Charm} we give the experimental quarkonia masses, the values of the parameter $M_0$ and our predictions for the masses of the excited states. It should be mentioned the the slope coefficient $\mu^2$ is universal for all trajectories and is defined according to the relation (\ref{eq:mucc}). The trajectories are shown on the figures \ref{fig:charm}c-g.

On the figure \ref{fig:charm}h we show the Regge trajectories on the $(J,M^2)$ plane for the ground states of charmonia
\beq
J &=& \alpha(M^2) =\alpha' M^2+\alpha(0). \label{eq:JM}
\eeq
Here we used the value of the slope coefficient from the proceeding analysis:
\beqa
\alpha' &=& \frac{1}{\mu_{c\bar c}^2},
\eeqa
and the interceptions $\alpha(0)$ for different trajectories were found to be
\beqa
\underline{J/\psi(1S), \chi_{c2}(1P)} &:& \alpha(0) = -2., \\
\underline{\eta_c(1S), \chi_{c1}(1S), h_c(1P)} &:& \alpha(0) = -2.8, \\
\underline{\chi_{c0}(1P), \psi(1D)} &:& \alpha(0) = -3.5.
\eeqa
These values are almost in the range $|\alpha_c(0)| \ge 3 \div 4$ that was obtained in the work with \cite{Khodjamirian:1992pz} the help of QCD sum rules. The  intercept of the parent trajectory was also presented in \cite{Kartvelishvili:1985ac}, where $\alpha(0)$ was expressed through the value of the $\JP$-meson at the origin. In that work the value $\alpha_c(0) = -3.5 \pm 0.6$ was found and it agrees well with the results of the present paper.

It is clearly seen that the positions of the physical states are described by the linear Regge trajectories with the  universal slope with a pretty good accuracy. It should be mentioned that for the leading Regge trajectory the exchange degeneration holds, i.e. $J/\psi$ and $\chi_{c2}$-mesons lie on one trajectory. This fact is not the result of the parameter fit, but automatic consequence of the correct description of the charmonia spectrum in $(n,M^2)$-plane. It should be mentioned that the exchange degeneracy was used earlier for the determination of the intercept $\alpha(0)$, that is important for determination of $c$-quark wave function in $J/\psi$, $\eta_c$, and $D$-mesons \cite{Kartvelishvili:1977pi,Kartvelishvili:1979bb,Kartvelishvili:1978id,Kartvelishvili:1978jh}.

\begin{table}
\input CharmTable
\caption{
The parameters of the Regge trajectories and masses of the mesons for $(c\bar c)$ sector. Our predictions for excited charmonium masses are shown in bold.
\label{tab:Charm}}
\end{table}

\section{New states}

Recently some new particles were observed. The production and decay channels of these particles indicate that they are excited charmonium states, so it would be interesting to consider the question of the disposition of this particles on the Regge trajectories presented in the previous section.

\subsection{$X(3872)$}

The first of such particles was the $X(3872)$-meson. The properties of this state were widely discussed in the literature \cite{Kalashnikova:2005ui,Swanson:2006st,Bauer:2005yu,Lesiak:2005ac}.

$X(3872)$ was observed by the Belle \cite{Choi:2003ue}, CDFII \cite{Acosta:2003zx}, D\O\ \cite{Abazov:2004kp} and BaBar \cite{Aubert:2004ns} collaborations in the decay
\beqa
B^\pm &\to& K^\pm X(3872) \to K^\pm \pi^+ \pi^- J/\psi,
\eeqa
and the world average of its mass is
\beqa
M_{X} &=& 3871.9 \pm 0.5 \MeV.
\eeqa
The further experimental study showed that besides the $X\to\pi^+\pi^- J/\psi$ decay the decays $X(3872)\to \gamma J/\psi$ \cite{Abe:2005ix}, $X(3872)\to\pi^+\pi^-\pi^0 J/\psi$ \cite{Abe:2005ix} \`e $X(3872)\to D^0\bar D^0\pi^0$ \cite{Bauer:2005yu} also exist and the branching fractions of these channels are linked by the relations
\beq
\Br(B^\pm\to K^\pm X) \Br(X\to \gamma \JP) &=& (1.8\pm 0.6 \pm 0.1)\times 10^{-6}, \nonumber\\
\Br(B^\pm\to K^\pm X) \Br(X\to \pi^+\pi^- \JP) &=& (1.3\pm 0.3 )\times 10^{-5}, \label{eq:BrABr2}\\
\Br(B^\pm\to K^\pm X) \Br(X\to D^0 \bar D^0 \pi^0) &=& (2.2\pm 0.7 \pm 0.4)\times 10^{-4},\nonumber \\
\frac{\Br(X\to \pi^+\pi^-\pi^0 \JP)}{\Br(X\to \pi^+\pi^-\JP)} &=& 1.0\pm0.4\pm 0.3.\nonumber
\eeq
For the total width of this state only the upper boundary is known: $\Gamma_X <2.3\MeV$. If we assume that there are no other significant decay modes, then we can determine the branching fractions of the reactions listed above and set the upper bounds on their widths:
\beq
\underline{X\to\gamma\JP} : \Br&\approx& (7.4\pm 0.4)\times 10^{-3},\qquad \Gamma < (17\pm 9)\keV,\nonumber\\
\underline{X\to\pi^+\pi^-\JP} : \Br&\approx& (5.3\pm 0.8)\%,\qquad \Gamma < (0.12\pm0.02)\MeV,\label{eq:Br2}\\
\underline{X\to\pi^+\pi^-\pi^0\JP} : \Br&\approx& (5.3\pm 4.6)\%,\qquad \Gamma < (0.12\pm0.11)\MeV,\nonumber\\
\underline{X\to D^0\bar D^0\pi^0} : \Br&\approx& (90\pm 2.5)\%,\qquad \Gamma < (2.1\pm0.6)\MeV.\nonumber
\eeq
Since the decay $X(3872)\to \gamma\JP$ is allowed, the charge parity of $X(3872)$ should be positive. The angular distribution in the $X(3872)\to\pi^+\pi^-\JP$ channel rules out the possibility of the scalar meson \cite{Abe:2005iy}.

Among the particles listed in the table \ref{tab:Charm} $\chi_{c0}(2P)$, $\chi_{c1}(2P)$ and $\eta_c(3S)$ mesons have masses that are most close to the mass of $X(3872)$. Since the case of the negative charge conjugation parity is forbidden by $X(3872)\to\gamma\JP$ decay and the case of the scalar  meson contradicts the angular distributions, the only variant left is $X(3872)=\chi_{c1}(2P)$.

There are, however, some arguments against this assignment. First of all, the upper bound for the width of the radiative decay $\Gamma(X(3872)\to \gamma \JP)<17\keV$ is less than theoretical predictions (for example, in \cite{Anisovich:2005vs} one can find the values $\Gamma[\chi_{c1}(2P)\to\gamma\JP]=30 \div 60\keV$). Second argument is that the decay $X(3872)\to\rho\JP\to\pi^+\pi^-\JP$ implies that $X(3872)$ is isovector, so it cannot be a charmonium. In the work \cite{Swanson:2006st} an alternative model is considered. According to this work $X(3872)$ is a loosely bound $D^0\bar D^{0*}$-molecule (deuson). The mass of $X(3872)$ is surprisingly close to $D^0 \bar D^{0*}$ threshold
\beqa
M_{D^0}+M_{D^{0*}} &=& 3871.2\pm 0.6\MeV
\eeqa
and in the case of zero orbital momentum of the mesons in this molecule its quantum numbers should be equal to $J^{PC}=1^{++}$. This assumption explains pretty good both the isospin violation and the smallness of the radiative decay width.

It should be mentioned that such an explanation have some serious drawbacks. First of all, the $X(3872)$ mass is above $D^0 \bar D^{0*}$ threshold. Secondly, the production probability of such a molecule should be smaller than the experimental value
\beqa
\Br(B^+ \to K^+ X) &=& (2.5\pm 1.0)\times 10^{-4}.
\eeqa
This value was obtained from the experimental results (\ref{eq:BrABr2}) and  (\ref{eq:Br2}) and has the same order of magnitude as the production probabilities of other charmonium states in similar decays \cite{Bauer:2005yu}. The smallness of the production rate  is caused by the fact that the size of the loosely bound molecule should be lager than that of the $(c\bar c)$ state, and the value of its wave function at the origin should therefore be smaller \cite{Suzuki:2005ha}. This problem could be avoided if we assume that $X(3872)$ is a mixture of the charmonium and $D\bar D^*$-molecule (in the weak decay of $B$-meson $(c\bar c)$ component of this mixture is produced, and the isospin violation in the decay $X(3872)\to\rho\JP$ is explained by the presence of $D D^*$-component), but the other partner of $X(3872)$ was never observed.

We think that the other explanation looks more plausible.
In the works \cite{Eichten:2004uh,Kalashnikova:2005ui} it was shown that closeness of $D\bar D^*$ thresholds (the mass of $X$ a little bit larger than the $D^0\bar D^{*0}$ threshold and a little bit smaller than the $D^+D^{*-}$ one) increases the probability of the mixing of $\chi_{c1}(2P)$ with other charmonium states. Such a mixing leads to the decrease of the radiative decay width and can be a reason for the isospin violation.

\subsection{$Z(3930)$, $X(3940)$, $Y(3940)$}

The state $Z(3930)$ was observed by the Belle collaboration in the reaction $\gamma\gamma \to D\bar D$ \cite{Abe:2005bp} and its mass equals
\beqa
M_{Z(3930)} &=& 3931 \pm 4 \pm 2\MeV.
\eeqa
The best candidates with the required charge parity from the particles listed in table \ref{tab:Charm} are $\chi_{c1}(2P)$, $\eta_c(3S)$ and $\chi_{c2}(2P)$ mesons. The production and decay channels  rule out  first two variants,  so we assign the quantum numbers $J^{PC}=2^{++}$ to $Z(3900)$.

The $X(3940)$ particle was observed by the Belle collaboration recoiling against $\JP$ in $\epem$ collision \cite{Abe:2005hd}. The mass of this particle is
\beqa
M_{X(3940)} &=& 3943\pm 6\pm 6\MeV
\eeqa
and the dominant decay chanel is $X(3940)\to D^* \bar D$. The best variants for this state are $\chi_{c1}(2P)$, $h_c(2P)$ and $\eta_c(3S)$ mesons. Production and decay channels exclude first two variants, so we choose the last assignment for $X(3940)$. It is interesting to mention, that our prediction for $\eta_c(3S)$ mass is only $19\MeV$ above the mass of $X(3940)$, while the predictions of other models are $4040-4060$, i.e. approximately $100\MeV$ too high \cite{Swanson:2006st}.

Belle has also observed one more particle in the region of 3940 MeV \cite{Abe:2004zs} --- $Y(3940)$. This meson was observed in the decay 
\beqa
B &\to& K Y(3940) \to K \omega \JP
\eeqa
and its mass is equal to 
\beqa
M_{Y(3940)} &=& 3943 \pm 11 \pm 13 \MeV.
\eeqa
The decays $Y(3940)\to D^{(*)}\bar D$ have not been seen, so it is possible that $X(3940)$ and $Y(3940)$ are dictinct states. Because of the existance of $Y(3940)\to\omega\JP$ the charge parity of $Y(3940)$ should be positive. The best variants for this particle from table \ref{tab:Charm} are $\chi_{cJ}(2P)$ mesons. Since the assigments for $\chi_{c1,2}(2P)$ are already chosen, we set $Y(3940)=\chi_{c0}(2P)$.

\section{Bottomonium}

\begin{figure}[h]
\includegraphics{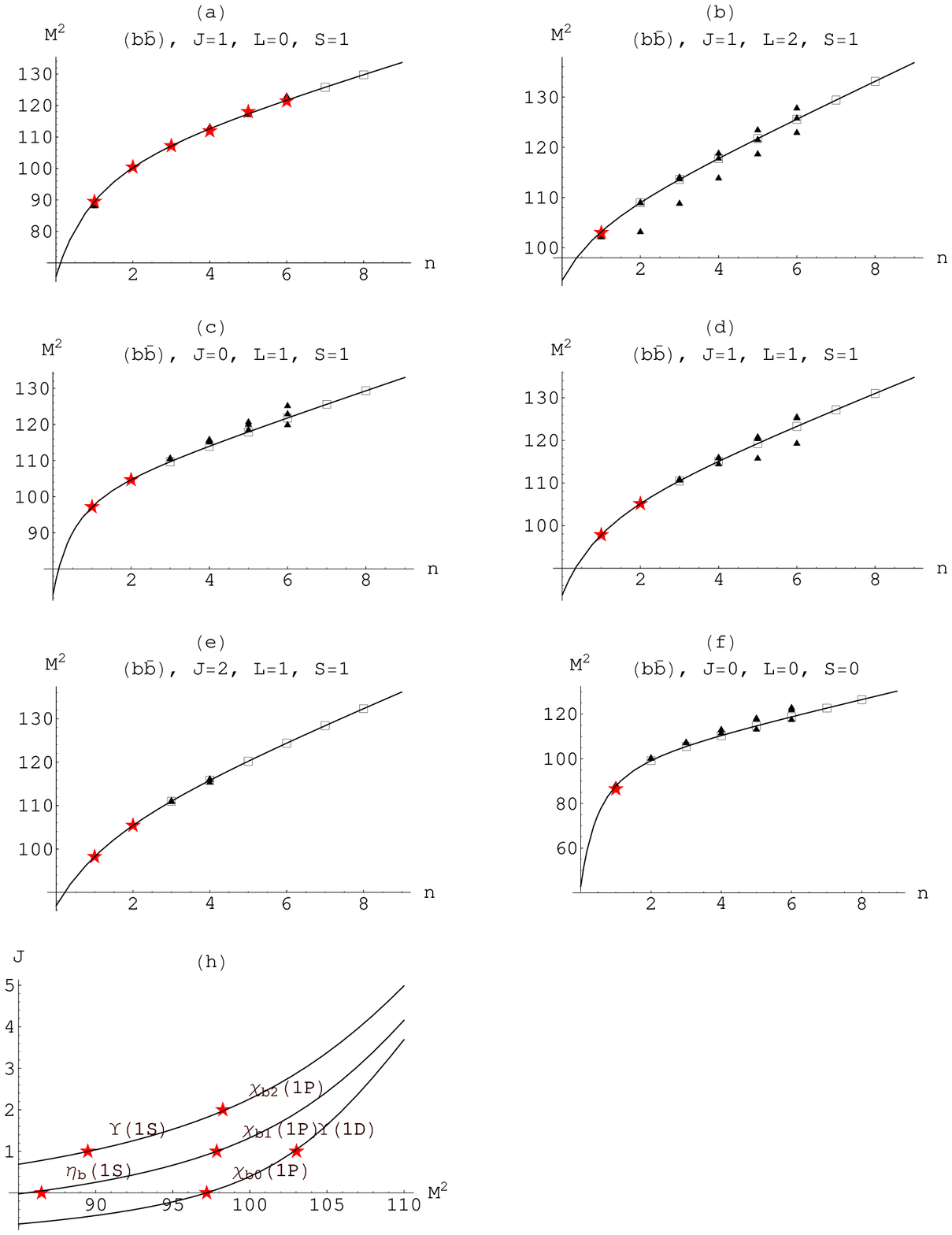}
\caption{
Regge trajectories for bottomonium states.%
$\bigstar$ --- experimental values, $\blacktriangle$ --- prediction presented in  \cite{Anisovich:2005vs},$\blacksquare$ --- the results of  \cite{Baldicchi:1998gt}, $\square$ --- our predictions
\label{fig:Bott}}
\end{figure}

In the previous sections we have shown that the charmonium states are described fairly good by the linear Regge trajectories on $(n,M^2)$ and $(M^2,J)$ planes with the slope coefficient
\beqa
\mu_{c\bar c}^2 &=& 3.2 \GeV^2, \qquad \alpha'=\frac{1}{\mu_{c\bar c}^2}.
\eeqa
The similar analysis for bottomonium states (i.e. the $(b\bar b)$ mesons) shows that in this case the linear  Regge trajectories describe the mass spectrum much worser. From figure \ref{fig:Bott}a, where all known vector bottomonia with the value of the orbital momentum $L=0$ are shown, one can see, that the slope of the Regge trajectory is not constant, but decreases with the increase of the radial quantum number $n$, so the linear Regge trajectory contradicts the experimental data. This fact should not be a big surprise. The linear Regge trajectories are typical for the "string" model of the quark-antiquark interaction under the assumption the the mass of the quarks tends to zero. Such an assumption is justified well for light mesons and agrees well with the experimental results. In the previous sections it was shown that the Regge trajectories of charmonium states are also linear, although the slope of these trajectories is different from that in the light meson sector. In the case of $(b\bar b)$ mesons, however, we see the significant deviations form the linearity. The possible reason of such deviation is that the massless quark approximations fails in this case. In the series of works (for example \cite{Batunin:1989gd,Ida:1977uy,LaCourse:1988cu,Pronko:1982nx}) the results obtained in the framework of the string model with massive quarks are presented and these results are in qualitative agreement with the experimental picture.

The same result can be obtained in the framework of the potential models. Let us consider the widely used Cornell potential of the quark-antiquark interaction \cite{Eichten:1978tg}:
\beq
V(r) &=& -\frac{4}{3}\frac{\alpha_s}{r} + \kappa r+c.
\label{eq:V}
\eeq
This potential combines the main characteristics of the quark interaction known from QCD. At small values of the distance between quark and antiquark $r\ll r_0\sim\sqrt{\alpha_s/\kappa}\sim 0.4\fm$ the leading term of this expression is the first one and we observe the asymptotical freedom. For large distances, on the contrary, the second term gives the main contribution and we observe the confinement. For the light mesons and charmonia the second case holds, so we can neglect the coloumbic term in the expression (\ref{eq:V}) and obtain the linear Regge trajectories on $(n,M^2)$ and $(J,M^2)$-planes
\beq
 M^2 &=& 8\kappa \left( 2n+J+\frac{3}{2}\right) + \mbox{const} \label{eq:kappa}
\eeq
that are in good agreement with the experimental data. For the lower bottomonium states, on the contrary, the first case takes place. After neglecting the second term in the expression (\ref{eq:V}) we get 
\beq
M^2 &=& -\frac{64\alpha_s^2 m^2}{9} \frac{1}{(n+J+1)^2}+\mbox{const}. \label{eq:coul}
\eeq
Linking the expressions (\ref{eq:kappa}) and  (\ref{eq:coul}) one can obtain the interpolation formula \cite{Sergeenko:1993sn}
\beq
M^2 &=&
  8\kappa \left( 2n+J+\frac{3}{2} \right)-
   \frac{b^2}{(n+J+1)^2}+M_0^2,
\label{eq:M2J}
\eeq
that joins both two limits. For large values of the quantum numbers the trajectory turns into linear with the slope
\beqa
\mu_{b\bar b}^2 &=& 16\kappa,
\eeqa
and for small values of the quantum numbers the spectrum of the nonrelativistic quarkonium is restored.

On the figures \ref{fig:Bott}a-f we show the Regge trajectories obtained with the help of the formula (\ref{eq:M2J}) for all bottomonium states. It should be mentioned that the slope coefficient 
\beqa
\mu_{b\bar b}^2 &=& 4.1\GeV^ 2
\eeqa
is universal for all this trajectories, while the parameters $b$ and $M_0$ depend on the spin and parities of the particle. It is clearly seen that the experimental states lie good on this trajectories and deviate from linear ones for small values of the radial quantum number $n$. The values of these parameters, as well as known experimental masses masses and the predictions obtained with the help of the formula  (\ref{eq:M2J}) are presented in the table \ref{tab:Bott}.

\begin{table}
\input BottTable
\caption{
Regge trajectories parameters and masses for $(b\bar b)$-mesons\label{tab:Bott}}.  Our predictions for excited bottomonium masses are shown in bold.
\end{table}

Up to now we have used the formula (\ref{eq:M2J}) for the construction of the Regge trajectories on $(n,M^2)$ plane and the spin of the particle for each trajectory was fixed. This formula can be used also for the construction of the trajectories on $(M^2,J)$ plane (Chew-Frautchi plot). To do this one needs to fix the radial quantum number $n$ and solve the equation (\ref{eq:M2J}) for the spin of the particle. There are three solutions of this equation, but only one of them is physically sensible (the others give negative values for the real part of $J$ and we will not consider them here). As it was shown earlier, in the case of charmonium states the parameters of the Regge trajectories on $(n,M^2)$ and $(M^2,J)$ planes coincide, so we have checked this property for bottomonia. The trajectories for lightest bottomonium mesons (i.e. $n=1$) are shown on figure \ref{fig:Bott}h and one can see that they agree well with the experimental data. 
One can observe that the exchange degenerations holds for the $1^{--}$ and $2^{++}$ states, as it was in $(c\bar c)$ case.
We did not perform any fits to obtain the values for the parameters of these trajectories. Instead of this we used the parameters for $S$ wave vector bottomonia (fig..\ref{fig:Bott}a) for the upper curve, and $1{++}$ and $0^{++}$ bottomonia (figures \ref{fig:Bott}d and \ref{fig:Bott}c) for the middle and lower curves respectively. In the limit of large $M^2$ all these trajectories become linear and the interceptions form them are equal to
\beqa
\underline{\Upsilon(1S), \chi_{b2}(1P)} &:& \alpha(0)=-0.88,\\
\underline{\eta_b(1S), \chi_{b1}(1P)} &:& \alpha(0)=-1.2,\\
\underline{\chi_{b0}(1P), \Upsilon(1D)} &:& \alpha(0)=-1.5.
\eeqa
These estimates, unfortunatelly, are not reliable. According to the QCD-sum rules restrictions \cite{Khodjamirian:1992pz} for the parent Regge trajectory we have
\beqa
\alpha_b(0) &<& -7 \div -8.
\eeqa
The value of the same order of magnitude was presented in \cite{Kartvelishvili:1985ac}, where the intersept $\alpha_b(0)$ was expressed though the leptonic width of the $\Upsilon(1S)$-meson:
\beqa
 \alpha_b(0) &=& -11.5 \pm 0.5.
\eeqa
There is also another argument against our values of the interceptions. It seems logical that the absolute value of the intercept of the corresponding Regge trajectory should be larger for larger quark masses.

In the work \cite{Brisudova:1999ut} other parametrization for Regge trajectories was used, and at first sight this parametrization is free from the mentioned drawback:
\beq
J &=& \alpha(M^2) = \alpha(0) + \gamma \left ( \sqrt{T}-\sqrt{T-M^2} \right), 
\label{eq:BrisA}
\eeq
where the "slope"
\beqa
 \gamma &=& 3.65 \pm 0.05 \GeV^{-1}
 \eeqa
 is universal and the values of $T$ and $\alpha(0)$ are
\beqa
\alpha(0) &=& -14.81 \pm 0.35, \\
\sqrt{T} &=& 12.50 \pm 0.02 \GeV.
\eeqa
It is clearly seen that the behaviour of this trajectory is completely different from the one considered in our paper. First of all, according to the expression (\ref{eq:BrisA}) the mass of the meson cannot exceed the threshold $\sqrt{T}$. The second difference is that the trajectory presented in \cite{Brisudova:1999ut} is linear for small meson masses and deviates from liear near the threshold, wihle the trajectory used in our pape is, on the contrary, linear for large $M^2$ and deiviates from linear for small.

The expression (\ref{eq:BrisA}) was obtained in the framework of the generalized string model, where the tension of the string depends on the quark separation:
\beqa
\sigma(r) &=& \frac{\sigma_0}{1+(\pi \mu r)^2},
\eeqa
and the parameters $T$, $\sigma_0$ and  $\mu$ are connected by the expression $T=\sigma_0/\mu$.  The same result can also be obtained in the framework of potential models if one use the potential
\beq
V_\mathrm{Br}(r) &=& \frac{\sigma_0}{\pi\mu} \arctan(\pi \mu r) \approx
 \left\{
   \begin{array}{ll}
    \sigma_0 r, & \quad r\ll 1/\mu,\\
    \sigma_0/(2\mu), & \quad r\gg 1/\mu.
   \end{array}
\right.
\label{eq:BrisV}
\eeq
The asymptotical behaviour of this potential is completely different from the behaviour if the cornell potential used in our work. For example for small values of quark separation  $V_\mathrm{Br}(r)$ is linear with respect to $r$, wile for cornell potential the coulomb term dominates. Thus, there is no assimptotic freedom in (\ref{eq:BrisV}).
Other difference is that for large $r$ the cornell potential increases infinitely, wihle $V_\mathrm{Br}(r)$ tendes to a finite constant. This could be tghe reason for the meson mass threshold existance in \cite{Brisudova:1999ut}.

\section{Conclusion}

The new states discovered recently in the $(c\bar c)$ sector open the question of their classification and the reliable prediction of heavy quarkonia masses. Such predictions were obtained in the framework of different potential or lattice models (see for example \cite{Anisovich:2005vs,Anisovich:2005jp,Chen:2000ej,Okamoto:2001jb,Ebert:2002pp,Barnes:2005pb}), but the results of this calculations depend strongly on the choice of the model parameters. For the mass of the $\chi_{c0}(2P)$ meson, for example, one can find in the literature the values from $3.822\GeV$ \cite{Anisovich:2005vs} to $4.080\GeV$ \cite{Chen:2000ej}. Since there is no reason to prefer some prediction it seems important to have some independent criterion that can help to choose the right value.

We think that this criterion could be the positioning of the mesons to the respective  Regge trajectories on $(n,M^2)$ and $(J,M^2)$ planes (here $n$ is the radial quantum number of the meson and $J$ and $M$ are its spin and mass). It is well known that the masses of the lights mesons can be described with a pretty good accuracy by the linear trajectories. In this paper we show, that this is, with minor changes, valid also for heavy quarkonia. Namely, the charmonia Regge trajectories are the straight line with the slope the is common for all charmonium mesons. In this paper we have used this trajectories to position to them recently discovered particles $X(3872)$, $Z(3930)$, $X(3940)$ and $Y(3940)$.

In the $(b\bar b)$ sector the situation slightly changes. For the small values of bottomonia masses we observe a deviation of the trajectories from linear, while for excited states linearity restores. This is the behavior that one should expect from different theoretical speculations, for example form the string picture of quark-antiquark interaction or potential models with Cornell potential. The results presented in our paper could be useful to improve the accuracy of $(b\bar b)$ masses predictions.

\begin{acknowledgments}
The authors thank professor G.P. Pronko  for useful discussions. We thanks Dr. S.Olsen, who pointed out some errors concerning our $Y(3940)$ assignment. This work was partially
supported by Russian Foundation of Basic Research under grant \#04-02-17530, Russian Education
Ministry grant \#E02-31-96, CRDF grant \#MO-011-0, Scientific School grant \#SS-1303.2003.2.
\end{acknowledgments}

\end{document}

%% file: CharmTable.tex
$$
\begin{array}{|c|c|c|c|c||c|c|c|c|c||}
\hline
 & n & \mbox{Data} & {M_0}, \GeV & {M_n}, \GeV &  & n & \mbox{Data} & {M_0}, \GeV & {M_n}, \GeV \\ 
\hline
 & 1 & 3.097 &  & 3.15\pm0.07 &  & 1 & 3.51 &  & 3.51\pm0.08 \\ 
 & 2 & 3.68609 &  & 3.63\pm0.09 &  & 2 &  &  & \mathbf{3.94\pm0.09} \\ 
 & 3 & 4.04 &  & 4.04\pm0.09 &  & 3 &  &  & \mathbf{4.33\pm0.1} \\ 
1^{--} & 4 & 4.415 & 2.6 & 4.42\pm0.1 & 1^{++} & 4 &  & 3.02 & \mathbf{4.68\pm0.1} \\ 
L=0 & 5 &  &  & \mathbf{4.77\pm0.1} & L=1 & 5 &  &  & \mathbf{5.01\pm0.1} \\ 
S=1 & 6 &  &  & \mathbf{5.09\pm0.1} & S=1 & 6 &  &  & \mathbf{5.32\pm0.1} \\ 
 & 7 &  &  & \mathbf{5.4\pm0.1} &  & 7 &  &  & \mathbf{5.61\pm0.1} \\ 
 & 8 &  &  & \mathbf{5.69\pm0.1} &  & 8 &  &  & \mathbf{5.89\pm0.1} \\ 
\hline
 & 1 & 3.77 &  & 3.76\pm0.09 &  & 1 & 3.556 &  & 3.56\pm0.08 \\ 
 & 2 & 4.16 &  & 4.17\pm0.1 &  & 2 &  &  & \mathbf{3.98\pm0.09} \\ 
 & 3 &  &  & \mathbf{4.53\pm0.1} &  & 3 &  &  & \mathbf{4.36\pm0.1} \\ 
1^{--} & 4 &  & 3.31 & \mathbf{4.87\pm0.1} & 2^{++} & 4 &  & 3.07 & \mathbf{4.72\pm0.1} \\ 
L=2 & 5 &  &  & \mathbf{5.19\pm0.1} & L=1 & 5 &  &  & \mathbf{5.04\pm0.1} \\ 
S=1 & 6 &  &  & \mathbf{5.49\pm0.1} & S=1 & 6 &  &  & \mathbf{5.35\pm0.1} \\ 
 & 7 &  &  & \mathbf{5.78\pm0.1} &  & 7 &  &  & \mathbf{5.64\pm0.1} \\ 
 & 8 &  &  & \mathbf{6.05\pm0.1} &  & 8 &  &  & \mathbf{5.92\pm0.1} \\ 
\hline
 & 1 & 3.415 &  & 3.42\pm0.08 &  & 1 & 2.979 &  & 3.05\pm0.07 \\ 
 & 2 &  &  & \mathbf{3.86\pm0.09} &  & 2 & 3.594 &  & 3.53\pm0.08 \\ 
 & 3 &  &  & \mathbf{4.25\pm0.1} &  & 3 &  &  & \mathbf{3.96\pm0.09} \\ 
0^{++} & 4 &  & 2.91 & \mathbf{4.61\pm0.1} & 0^{-+} & 4 &  & 2.47 & \mathbf{4.35\pm0.1} \\ 
L=1 & 5 &  &  & \mathbf{4.95\pm0.1} & L=0 & 5 &  &  & \mathbf{4.7\pm0.1} \\ 
S=1 & 6 &  &  & \mathbf{5.26\pm0.1} & S=0 & 6 &  &  & \mathbf{5.03\pm0.1} \\ 
 & 7 &  &  & \mathbf{5.56\pm0.1} &  & 7 &  &  & \mathbf{5.34\pm0.1} \\ 
 & 8 &  &  & \mathbf{5.84\pm0.1} &  & 8 &  &  & \mathbf{5.63\pm0.1} \\ 
\hline
\end{array}
$$

%% file: BottTable.tex
$$
\begin{array}{|c|c|c|c|c||c|c|c|c|c||}
\hline
 & n & \mbox{Data} & {M_0}, \GeV & {M_n}, \GeV &  & n & \mbox{Data} & {M_0}, \GeV & {M_n}, \GeV \\ 
\hline
 & 1 & 9.46 &  & 9.46\pm0.1 &  & 1 & 9.892 &  & 9.89\pm0.1 \\ 
 & 2 & 10.023 &  & 10.\pm0.1 &  & 2 & 10.255 &  & 10.3\pm0.1 \\ 
 & 3 & 10.355 &  & 10.4\pm0.1 &  & 3 &  &  & \mathbf{10.5\pm0.1} \\ 
1^{--} & 4 & 10.58 & b=12.1 & 10.6\pm0.1 & 1^{++} & 4 &  & b=8.68 & \mathbf{10.7\pm0.1} \\ 
L=0 & 5 & 10.865 & {M_0}=9.87 & 10.8\pm0.1 & L=1 & 5 &  & {M_0}=9.9 & \mathbf{10.9\pm0.1} \\ 
S=1 & 6 & 11.02 &  & 11.\pm0.1 & S=1 & 6 &  &  & \mathbf{11.1\pm0.1} \\ 
 & 7 &  &  & \mathbf{11.2\pm0.1} &  & 7 &  &  & \mathbf{11.3\pm0.1} \\ 
 & 8 &  &  & \mathbf{11.4\pm0.2} &  & 8 &  &  & \mathbf{11.4\pm0.2} \\ 
\hline
 & 1 & 10.15 &  & 10.2\pm0.1 &  & 1 & 9.912 &  & 9.91\pm0.1 \\ 
 & 2 &  &  & \mathbf{10.4\pm0.1} &  & 2 & 10.268 &  & 10.3\pm0.1 \\ 
 & 3 &  &  & \mathbf{10.7\pm0.1} &  & 3 &  &  & \mathbf{10.5\pm0.1} \\ 
1^{--} & 4 &  & b=6.58 & \mathbf{10.9\pm0.1} & 2^{++} & 4 &  & b=12.5 & \mathbf{10.8\pm0.1} \\ 
L=2 & 5 &  & {M_0}=9.99 & \mathbf{11.\pm0.1} & L=1 & 5 &  & {M_0}=9.9 & \mathbf{11.\pm0.1} \\ 
S=1 & 6 &  &  & \mathbf{11.2\pm0.1} & S=1 & 6 &  &  & \mathbf{11.2\pm0.1} \\ 
 & 7 &  &  & \mathbf{11.4\pm0.2} &  & 7 &  &  & \mathbf{11.3\pm0.1} \\ 
 & 8 &  &  & \mathbf{11.5\pm0.2} &  & 8 &  &  & \mathbf{11.5\pm0.2} \\ 
\hline
 & 1 & 9.859 &  & 9.86\pm0.1 &  & 1 & 9.3 &  & 9.37\pm0.1 \\ 
 & 2 & 10.232 &  & 10.2\pm0.1 &  & 2 &  &  & \mathbf{9.96\pm0.1} \\ 
 & 3 &  &  & \mathbf{10.5\pm0.1} &  & 3 &  &  & \mathbf{10.3\pm0.1} \\ 
0^{++} & 4 &  & b=5.26 & \mathbf{10.7\pm0.1} & 0^{-+} & 4 &  & b=7.43 & \mathbf{10.5\pm0.1} \\ 
L=1 & 5 &  & {M_0}=9.89 & \mathbf{10.9\pm0.1} & L=0 & 5 &  & {M_0}=9.76 & \mathbf{10.7\pm0.1} \\ 
S=1 & 6 &  &  & \mathbf{11.\pm0.1} & S=0 & 6 &  &  & \mathbf{10.9\pm0.1} \\ 
 & 7 &  &  & \mathbf{11.2\pm0.1} &  & 7 &  &  & \mathbf{11.1\pm0.1} \\ 
 & 8 &  &  & \mathbf{11.4\pm0.2} &  & 8 &  &  & \mathbf{11.2\pm0.1} \\ 
\hline
\end{array}
$$